# Climate change adaptation stories: co-creating climate services with reindeer herders in Finland

Marta Terrado[1], Nuria Pérez-Zanón[1], Dragana Bojovic[1], Nube González-Reviriego[1], Gerrit Versteeg[1], Sara Octenjak[1], Albert Martínez-Botí[1], Tanja Joona[2]

[1] Barcelona Supercomputing Center (BSC-CNS), Plaça Eusebi Güell, 1-3, 08034 Barcelona, Spain

[2] Arctic Centre, University of Lapland, Pohjoisranta 4, 96100 Rovaniemi, Finland

[*] corresponding author: Marta Terrado, marta.terrado@bsc.es

## Abstract

Reindeer husbandry in the Arctic region is strongly affected by the local climate. Reindeer herders are used to coping with adverse weather, climate, and grazing conditions through autonomous adaptation. However, today's rapidly changing Arctic environment poses new challenges to the management of herding activities. Finding means for combining traditional and scientific knowledge without depriving any of the systems of its fundamental strengths is hence deemed necessary. In this work, we apply a transdisciplinary framework for knowledge co-production involving international researchers and reindeer herders from different cooperatives in northern Finland. Through 'climate change adaptation stories', we co-explore how climate predictions can inform herders' decision making during the herding season. Relevant decisions include the anticipation of summer harvest time, the inopportune periods of cold weather in spring, and insect harassment in summer. Despite their potential benefits for climate-sensitive decisions, climate predictions have seen limited uptake, mainly due to their probabilistic nature and lower quality compared with shorter-term weather forecasts. The analysis of two different adaptation stories shows that seasonal predictions of temperature for May and June can successfully advise about the likelihood of having an earlier than normal harvest. This information can be obtained up to three months in advance, helping herders to better arrange their time for other activities. Likewise, sub-seasonal predictions of temperature during April and May can be useful to anticipate the occurrence of backwinter episodes, which can support herders in deciding whether to feed reindeer in pens for longer, avoiding putting the survival of calves at risk. This study, which would benefit from co-evaluation in real world settings and consideration of additional adaptation stories, sets the basis for a successful co-production of climate services with Arctic reindeer herders. This research shows the potential to enhance the resilience of Polar regions, offering opportunities for adaptation while supporting the sustainability and culture of traditional practices of Arctic communities.

## 1. Introduction

Reindeer husbandry is practised in the regions of Fennoscandinavia and Kola Peninsula by more than 20 different ethnic groups (CAFF 2006). Decision making within a reindeer herding community has been based on traditional knowledge (encompassing indigenous and local practitioner knowledge), which are contextual and local forms of knowledge passed on from one generation to the next (Rasmus et al. 2020). Despite the capacity of reindeer herders to cope with adverse weather and grazing conditions through autonomous adaptation, also referred to as resonance strategies (Malik et al. 2010, Vuojala-Magga et al. 2011), today's rapidly changing Arctic environment poses new challenges to the management of herding activities.

Indeed, changes in climate are occurring significantly faster in the Arctic than in the rest of the globe (Smith et al. 2021). According to the Sixth IPCC Assessment Report (Constable et al. 2022), increasing heatwaves, wildfires, extreme precipitation, permafrost loss and rapid seasonal snow and ice thaw events will threaten terrestrial subsistence food resources across the Arctic. Reindeer husbandry is affected throughout the herding year by the local climate, either directly through the effects of weather events on the well-being of reindeer or on herding work, or indirectly through the impact of weather conditions on pastures (e.g. Heggberget et al. 2002, Oskal 2008, Oskal et al. 2009, Turunen et al. 2016, Eira et al. 2013, 2018). The increased weather variability and unpredictability of weather and climate may hinder the use of traditional knowledge as that knowledge may no longer correspond with the changing conditions (Axelsson-Linkowski et al. 2020, Turunen et al. 2021). In the case of Finland, a vast body of literature has focused on the impacts of a changing climate and difficult weather and snow conditions on reindeer husbandry, documenting strategies applied by herders to cope with these conditions (Heikkinen et al. 2012, Turunen et al. 2016, Rasmus et al. 2018, 2020, Mathiesen et al. 2023). A number of these studies have focused on winter, which is a critical season for herding because difficult snow conditions can decrease the availability of winter forage (Turunen et al. 2016, Heikkinen et al. 2012, Rasmus et al. 2018). The more frequent frost and thaw cycles, added to autumn precipitation, leads to deep snow or ice formation in the snowpack that causes reindeer to spend more energy digging and moving, which can affect animal condition (Tyler et al. 2007, Helle and Kojola 2008). Adding to that, a shorter snow season with later snow cover formation and earlier snow melting has been described elsewhere (Turunen et al. 2016, Lépy and Pasanen 2017, Luomaranta et al. 2019). However, Forbes and co-authors reported about a workshop organised in Finland, where reindeer herders urged scientists to also pay attention to summer pastures (Forbes et al. 2006). They felt that the emphasis on winter pastures in recent years was at the expense of understanding summer pastures, which are important for calf growth and overall herd production (Kumpula et al. 1999).

With the aim to be more comprehensive and cover the entire herding year, Rasmus and co-authors (2020) explored the interannual variability and changes over time in temperature, precipitation and snow related climate indices (annual and seasonal) relevant for reindeer husbandry in northern Finland. The authors combined observations of weather data from the past 30 years with reindeer herder knowledge (experiences and perceptions) collected through a questionnaire. Relying on herders' perceptions connects to the concept of episodic memory (i.e. reliving events), which in the climatic literature has been identified as effective to make risks more tangible. As a result, users are more prone to respond to these risks since they have experienced them (Shepherd et al. 2018). Another study from Turunen et al. (2021) analysed the experiences and perceptions of coping with cold among physically active herders through semi-structured interviews. They concluded that coping with extreme conditions not only requires flexibility, preparedness, and innovation from the herders but also thoughtful caution when approaching and managing unexpected situations.

Studies like the ones described above provide evidence of how necessary observations are to better understand and monitor the changes in weather and climate conditions as well as learn how to deal with these changes. However, forecasts also have a pivotal role in adapting to climate change, as it involves being prepared for the future. Besides studies that focus on climate projections for the prediction of longer-term conditions (e.g. Rees et al. 2006, Tyler et al. 2007, 2021, Vistnes et al. 2009), to our knowledge this is the first study that explores the application of climate predictions at sub-seasonal and seasonal time-scales for reindeer husbandry.

Sub-seasonal-to-seasonal (S2S) climate predictions try to anticipate the most likely climate conditions for the next few weeks and months into the future (Doblas-Reyes et al. 2013, Manrique-Suñén et al. 2023). These predictions, which sit between weather forecasts (for the next few hours and days) and decadal predictions (for the next few years), have the potential to inform different climate-vulnerable sectors in adapting their short- to medium-term strategies to climate variability and change (e.g. Lowe et al. 2017, Lledó et al. 2019, Manrique-Suñén et al. 2023, Vigo et al., 2023). However, despite their potential benefits, there has been limited uptake of climate predictions by users (White et al. 2017). Major limitations relate to the difficulty to deal with probabilistic climate information and their lower skill (i.e. quality of the prediction based on its performance in the past) compared with the skill of weather forecasts, among other factors (Terrado et al. 2019). However, under a changing climate, there is indeed potential for S2S predictions to help anticipate adverse and favourable climatic conditions, including extreme events and their impacts.

Due to the different but complementary nature of scientific and traditional knowledge, finding means for combining both knowledge systems is instrumental to increase the resilience of reindeer herders to climate change (Fenge and Funston 2009). This can be achieved through the application of transdisciplinary co-production frameworks, understood as iterative, interactive and collaborative processes that bring together a plurality of knowledge sources to mutually define problems and develop usable products to address these problems (e.g., Armitage et al. 2011, Moore and Hauser 2019, Nörstrom 2020, Bojovic et al. 2021, Terrado et al. 2023a). In the context of this study, usable products refer to products that are understandable, quality-assured, and fit for supporting herders' decision-making (e.g. the likelihood of having an inopportune episode of cold weather in Spring).

People and knowledge systems brought together in knowledge co-production processes may come from different backgrounds and experiences, often connecting academics and non-academics (Baulenas et al. 2023). However, the application of co-production frameworks that specifically seek to equitably bring together people from different knowledge systems (scientific and traditional knowledge) are less common, although they have seen an increase in the literature (e.g., Thornton and Maciejewski Scheer 2012, van Bavel et al. 2020, Wheeler et al. 2020, Yua et al. 2022). Efforts to bring local and traditional knowledge into global reports on climate change are also emerging, such as the IPCC report, as noticed by Mustonen et al. (2021). In the context of reindeer husbandry, co-production approaches have also been applied in response to social-ecological changes (Pape and Löffler 2012, Sarkki 2013, Forbes et al. 2019, Horstkotte et al. 2021, Krarup Hansen et al. 2022), with various works focusing on changes in climate and land use (Maynard et al. 2011, Rasmus et al. 2020).

In this study, we apply a transdisciplinary framework for knowledge co-production, involving researchers and reindeer herders from various cooperatives in northern Finland. Through so-called 'adaptation stories', we explore with reindeer herders how future S2S climate predictions can inform their planning and management practices, considering that climate change will continue to greatly modify herders' working conditions in the future. Stories can describe herders' experiences, but also facts, contexts and imaginaries. In this work, we also use stories to increase the herders' understanding of and engagement with climate science, as claimed by Bloomfield and Manktelow (2021). This approach has been previously applied to co-produce knowledge with Arctic communities through climate related case studies (Terrado et al. 2023b), which were identified as a useful tool to add granularity to the challenge of climate change adaptation.

The objective of our study is twofold, focusing on the co-production process rather than the specific resulting products, which tend to be case-specific. First, we aim to co-explore with a group of reindeer herders the usefulness of S2S predictions, a source of information that has traditionally not been considered in herders' decision making, and co-develop climate adaptation stories which can serve as the basis for future climate services. This is done through participatory activities including informal interviews, a survey and a roundtable. Climate adaptation stories are used here as a means to build capacity, ensuring that reindeer herders understand the opportunities of probabilistic climate information, but also the limitations. By doing so, we also attain our second objective, which is to understand how scientific knowledge on climate can complement the traditional knowledge of reindeer herders providing additional opportunities for smarter adaptation to climate change. Integrating knowledge systems is important to move from comprehensive science to useful knowledge and services.

## 2. Background information

### 2.1. Case study context: reindeer husbandry in northern Finland

In Finland, reindeer husbandry is a traditional livelihood for indigenous Sámi, as well as non-indigenous herders. The reindeer management area (RMA) in northern Finland is divided into 54 herding cooperatives called *paliskunta* (https://paliskunnat.fi/map/), each responsible for the proper management of reindeer in its own area (Figure 1). All herding cooperatives belong to the Reindeer Herders' Association of Finland. Every reindeer owner is a member of the local association, although the reindeer are owned individually. However, in the Sámi area in northernmost Lapland some of the herding cooperatives follow their own traditional system, in which the reindeer are managed by smaller collective units called *siidas*.

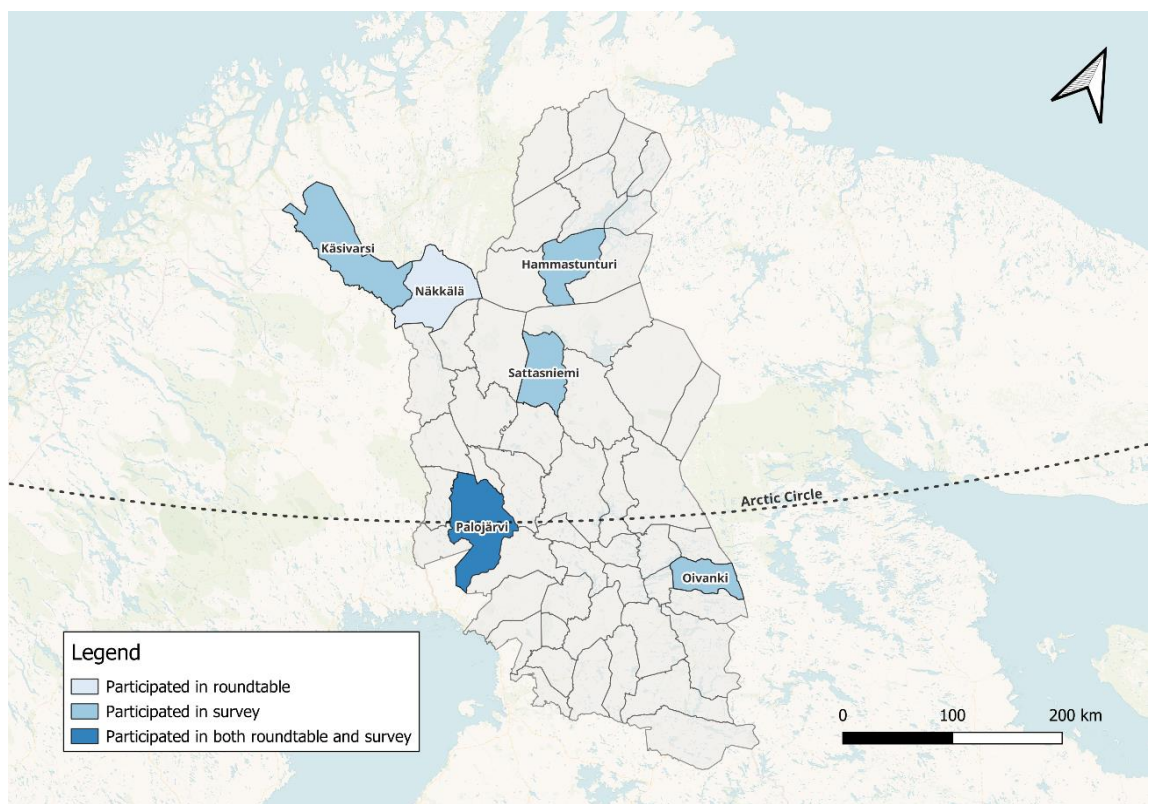

**Figure 1:** Cooperatives in the reindeer management area in northern Finland. Coloured cooperatives highlight those whose herders participated in the survey and/or roundtable.

The reindeer management area can be divided into three main regions: 'the Sámi homeland' (northern region, dominated by fell with tundra), which officially also belongs to the 'area specifically intended for reindeer management' (middle region, dominated by coniferous tree forests and where the biggest mining developments are located). Finally, there is 'the reindeer herding area' (southern region, covered by boreal forests in its northernmost part and bog plains in its southern part), where various industries are located (Reindeer Herders' Association 2014, Horstkotte et al. 2021, Landauer et al. 2021).

There are various differences between herding practices and climate and environmental conditions in cooperatives located in the northern and southern parts of the reindeer management area in northern Finland. These differences between northern mountainous districts and southern forested districts have been reported in the literature (Vuojala-Magga et al. 2011, Turunen and Vuojala-Magga 2014, Rasmus et al. 2020). Some relevant aspects include that feeding in pens during the coldest winter months is more frequent in the southern areas while field feeding is a more usual practice in the northern and central areas (Horstkotte et al. 2020).

The herding year is defined as the period from the 1st of June to the 31st of May of the next year. A detailed description of the seasonal cycle in reindeer life, pasture use, and herding can be found in the literature (Heikkinen et al. 2012, Turunen and Vuojala-Magga 2014, Turunen et al. 2016, 2021, Rasmus et al. 2020). From mid-summer onwards, the new calves are earmarked with the owner's reindeer earmark (Reindeer Husbandry Act, 848/1990). The herding work in the summer includes haymaking to be prepared for possible supplementary feeding during the following winter. In the fall and winter, herders' work in northern Finland includes collecting and moving reindeer herds to round-up sites, working in round-ups, slaughtering and processing meat as well as daily feeding and monitoring of the animals in the field. Therefore, working days during this period are the longest and the work is most physically demanding (Turunen et al. 2021). During the winter, reindeer prefer to feed on ground-growing lichens, which they dig through the snowpack, although arboreal lichens can be an important resource in the late winter (Heggberget et al. 2002, Helle and Jaakkola 2008, Turunen et al. 2009, Ophof et al. 2013). Reindeer are also given supplementary forage in the field or in enclosures if needed, especially in late winter or during difficult snow conditions (Horstkotte et al. 2021). At the end of the herding year, new calves are born either in open-range or enclosure calving. In open-range calving, reindeer give birth to their calves in their natural calving grounds, whereas in enclosure calving, pregnant reindeer are moved into a separate section within the enclosure from March–April until the end of May. Enclosure calving allows herders to have a higher control of the herd and ensure calves' growth, helping the animals to deal with extreme weather conditions, providing them with extra nutrition and protection from predators (Rasmus et al. 2021).

## 2.2. Climate predictions

In this work, a 'climate forecast product' is defined as climate information tailored to the users' context (e.g. the likelihood of having an inopportune period of cold weather next Spring). A climate forecast product is built from the output of a climate forecast system, which results from running the system for several time steps in the future, given a set of initial conditions. The length of the forecast runs (i.e. the number of time steps in the future), the specific initialization date (e.g. 1st of January 2020), the number of perturbed initial conditions (i.e. number of ensemble members) and the atmospheric variables saved (e.g. surface air temperature) define the climate forecast output. These outputs can later be post-processed in several ways (WMO 2019). These include, for instance, time averages, spatial aggregations on the area of interest, selection of periods and initialization dates, as well as reduction of the systematic model errors by applying specific bias-adjustment methodologies (Pérez-Zanón et al. 2022). Information on the context and the climate forecast application is crucial to decide on the best post-processing methodology for a particular climate forecast product.

A climate forecast system differs from a weather forecast system in that, apart from the modelisation of the atmosphere, it also includes the oceans, sea ice, and/or land components as well as their interaction. These components become necessary when the forecast length (time distance from the initial conditions) increases because the information from the initial conditions fades, and these components bring other sources of predictability. For example, the evolution of the sea surface temperatures plays a key role in some regions of the Earth, and periodic phenomena such as El Niño Southern Oscillation (ENSO) in the Pacific can affect conditions in different places around the globe (Davey et al. 2014). However, not all atmospheric variables are affected in the same way. While accumulated precipitation in a specific season and region may increase its predictability, the temperature is not necessarily affected in the same way.

Given the described limitations on predictability, it is not always possible to provide climate forecasts that fulfil user's requests. For instance, some of these requests are for short-term events (e.g. ice formation on the ground), which are different from longer-term climate events (Diaz and Murnane 2008), because they depend on several constraints that relate to the field of meteorology and that fall outside of the climate science field.

## 3. Knowledge co-production framework

We applied elements of a co-production approach developed for climate services projects that includes the steps of co-exploration, co-design and co-development (Bojovic et al. 2021, Terrado et al. 2023b). This framework encourages transdisciplinary processes, i.e., interaction between scientists and stakeholders. The prerequisite is an interdisciplinary collaboration (Bojovic et al. 2021), which in this work occurred between academic partners, including natural scientists, social scientists and science communicators. This interdisciplinary collaboration helped the selection of appropriate climate data, combining it with other types of local data, and the transformation of this data into climate information and services (see sections 3.3 and 3.4).  Adding other types of knowledge to this interdisciplinary collaboration was important to move from comprehensive science to useful knowledge and services. A transdisciplinary approach, which is the key for a successful co-production, was hence ensured in parallel by the participation of a research partner that is also a herder of the Palojärvi reindeer herding cooperative. The participation of this partner was key for engaging with other members of the cooperatives in northern Finland. Namely, this partner provided a link to and helped build relationships with reindeer herders who took an active role in the identification of the research questions through co-exploration of the key weather conditions that affect herding and actions used in response to them. It also allowed a more precise understanding of the herders' attitudes towards climatic risks and a more accurate interpretation of the answers gathered through the survey, informal interviews, and the roundtable discussion (see sections 3.1 and 3.2). This knowledge exchange was used for the co-design of climate adaptation stories. Finally, this close collaboration with a sample of reindeer herders led to the co-development of new climate knowledge that could be the basis for future climate services.

### 3.1. Informal interviews with the Palojärvi reindeer herder

Monthly meetings in the form of informal interviews were held involving the transdisciplinary team. These interviews were used to discuss different issues with the reindeer herder from the Palojärvi cooperative. Topics ranged from specificities and main challenges at different moments of the herding year, to how to best engage with other reindeer herders, and how factors other than climate affect the herding activity. These conversations helped the team to have a common understanding of the herders' context in northern Finland to better define the characteristics for a future climate service.

### 3.2. Survey on reindeer husbandry and climate change

A survey was conducted with reindeer herders from northern Finland to assess how practitioners' knowledge was integrated in day-to-day decisions and to identify how the forecasts of weather and climate conditions could help communities adapt to climate change. Participants were given a link to the online survey created using Survey Monkey. The survey was anonymous but included introductory questions about the participant's cooperative, age group and gender, followed by questions about the key weather conditions that affect reindeer herding in different seasons. A list of key weather conditions was identified based on a literature review and preliminary informal interviews with reindeer herders. The survey guided participants to select the weather conditions observed in each season that had been more relevant for them in the past years. In order to explore into more details the particular weather conditions indicated by users, the selected choice would  trigger the survey branching logic, i.e. an individualised set of sub-questions that correspond to the selected weather conditions would open for the user. The sub-questions inquired the following: how the weather condition affected reindeer herding, the years when this condition was observed, how herders addressed the impact, and finally, what climate information could help them adapt to the selected condition and how much in advance should this information be provided to be useful. The following set of questions inquired herders about their current sources of weather and climate knowledge and how this information is usually received and shared. Another set of questions focused on the main factors and risks that affected participant's activities. For that, we provided a list of risks that participants had to select from, which included different types of conflicts, land management, geopolitical situation, and environmental changes. This was followed by an open-ended question about how participants addressed these risks. To finish, the survey asked about any significant recent changes observed in reindeer behaviour and reindeer herding practice. Most of the questions were closed-ended questions with the possibility to leave additional information. The list of survey questions can be found in the Supplementary Material (Text S1). The online survey was launched at the beginning of June 2022 and was open for 3 months, until the end of August. It was initially developed in English and then translated to Finnish. The link to the Finnish version of the survey was shared on social media (Facebook from the Arctic Centre in the University of Lapland and the Finnish Association of Reindeer Herders) and through in-person contact. Answers from 11 reindeer herders were collected in total. Survey results aimed at confirming which findings from those available in the literature were also applicable to the context of the cooperatives in Finnish Lapland, while identifying novel applications of climate information which are case specific. Although the number of answers was low, it was sufficient in the context of this study as they covered a good range of extreme events and coping strategies  that could be taken as a starting point to illustrate the potential application of climate predictions in the context of reindeer herding.

### 3.3. Roundtable with reindeer herders

After analysing the survey results, we organised a roundtable discussion with a group of reindeer herders. The aim of the roundtable was to assess the herders' response to the preliminarily identified weather-related stories, define appropriate climate service products, and share examples of the type of information that the project would be able to deliver. The roundtable structure was streamlined in a presentation format, showcasing each season's crafted stories. The slides included the reasoning behind the story, which was related to a particular herder's decision making point, as well as climate observations and predictions for years of interest, a quality assessment of the prediction, and a list of discussion points to clarify some survey outputs. This information was initially prepared in English and then translated to Finnish. The roundtable discussion took place on the 4th of January 2023, in Näkkälä with participants from both the Näkkälä and Palojärvi reindeer herding cooperatives. Näkkälä belongs to Enontekiö's municipality and is situated in the north of Finland, where the landscape is dominated by barren mountainous features. The reindeer herder partner facilitated the roundtable discussion, which was conducted in Finnish and involved seven herders, 2 women and 5 men aged between 40 and 63 years old, all with extensive experience in reindeer herding.

### 3.4. Selection of climate data

Two climate prediction systems have been explored to provide information for the climate adaptation stories: the sub-seasonal climate prediction from the National Centers for Environmental Predictions Climate Forecast System (NCEP CFSv2; Saha et al. 2014) and the seasonal climate forecast from the European Centre for Medium-Range Weather Forecasts (ECMWF) System 5 (SEAS5; Johnson et al. 2019). These climate forecasts include additional sources of predictability than the initial conditions to capture impacts from the large-scale circulation patterns for the next weeks (sub-seasonal) and months (seasonal). The information on surface climate variables at sub-seasonal and seasonal time scales is aggregated in time periods of weeks and months, respectively, and has a probabilistic nature (probabilities instead of exact values) to account for model uncertainty.

Retrospective sets of forecasts for the past, called hindcasts, are produced for each prediction system and are used to correct systematic errors and calculate metrics to evaluate the performance of the prediction systems. Due to computational constraints, the hindcasts usually have fewer ensemble members or initializations than the real-time forecasts. An ensemble member is a set of forecasts (rather than a single one) generated from running a computer model a number of times from slightly different starting conditions.

The sub-seasonal forecast NCEP CFSv2 is a fully coupled system representing the interaction between the Earth's atmosphere, oceans, land and sea ice. The forecast length for sub-seasonal predictions is 45 days and the system is initialised every 6 h (00, 06, 12 and 18 UTC). The hindcast period is 1999-2018 with one ensemble member for each 6-hourly initialization. The forecast period, which started in 2019, has four ensemble members for each 6-hourly initialization (3 perturbed members and one control run). The model output is then processed to produce a lagged ensemble by combining the model runs of three days. This method produces an ensemble of 12 ensemble members for the hindcast period and 48 members for the forecast. The raw data from NCEP CFS v2 is freely available from the NCEP NOMADS server at a resolution of 1º (~100km). To maximise the use of these datasets, the period 1999-2021 is used by combining initialization from hindcast and forecasts. For the forecast period, only 12 members were selected.

The seasonal forecast SEAS5 is a coupled system of ocean and sea ice. The forecast consists of a 51-member ensemble initialised every month on the first day of the month and integrated for seven months. The hindcast of seasonal predictions covers 1981-2017 and consists of 25 ensemble members. The raw data from SEAS5 is freely available from the Copernicus Climate Change Service (C3S) Climate Data Store (CDS) at the resolution of 1°.

As mentioned, these climate predictions can be post-processed and assessed by comparing them against reference datasets. Two reference datasets were initially considered: the ERA5-Land (Muñoz-Sabater et al. 2021) and the observational gridded dataset from the Finnish Meteorological Institute (FMI; Aalto et al. 2016). Comparing the spatial monthly climatologies from both datasets for northern Finland showed a smoothness of extreme values in some months from ERA5-Land (Figure S1 in the Supplementary Material). On the other hand, the FMI gridded dataset spatially interpolates in-situ observations from weather stations at a higher resolution (1 km$^2$). Therefore, it was more suitable for developing a product tailored to the needs of reindeer herders in the context of this study.

### 3.5. Climate information and services

To get an understanding of the magnitude and length of climatic events affecting the reindeer herding activity, and therefore selecting the most suitable S2S forecasting product, we used the PlotWeeklyClim function developed for CSTools (Pérez-Zanón et al. 2022). This tool allowed us to compare the weekly and daily values of climate variables, aggregated for the whole reindeer management area in northern Finland, against the weekly climatological values for the dates when herders observed the event. This comparison allowed us to see whether herders' perceptions were reflected in the observational records.

For spatial representation, climate data were aggregated at the municipality level using the Local Administrative Units provided by the Eurostat database (https://ec.europa.eu/eurostat/web/nuts/local-administrative-units), which are compatible with the nomenclature of territorial units for statistics (NUTS). Since the reindeer management area in Finland is divided in herding districts, the limitations of using the municipal division were discussed with reindeer herders, who indicated that weather information in Finland is provided at the level of municipality, and therefore, such information is sufficient. The second step consisted in calculating the climatological tercile categories and extreme percentiles (Wilks 2011) for each municipality in the reindeer management area. The category in which the target event fell was determined considering five categories: (i) lower extreme (values below 10th percentile; p10), (ii) below normal (values between 10th and 33rd percentile; p10 - p33), (iii) normal (values between 33rd and 66th percentiles; p33 - p66), (iv) above normal (values between 66th and 90th percentile; p66 - p90), and (v) upper extreme (values above 90th percentile; p90).

Finally, the selected climate predictions (either sub-seasonal or seasonal) were post-processed using the observed event as a target date to evaluate if the predictions could anticipate it. Post-processing was applied to improve the applicability of climate data to the context under study. Seasonal climate predictions were bias-adjusted using the method described in Torralba et al. (2017), whereas for sub-seasonal predictions the methodology described in Manrique et al. (2023) was used to take advantage of more than one initialization per week, which allowed us to better correct the biases. Then, the fair Ranked Probability Skill Score (fRPSS; Wilks 2011) on tercile categories was calculated using the climatology as a reference forecast. fRPSS values below zero are defined as unskilful, those equal to zero indicate that the forecast provides similar information to the climatology, and fRPSS greater than zero shows that the predictions are skilful, which means they are better suited for decision making than taking the climatology (i.e. the average climate conditions of the past) as reference for future conditions.

## 4. Results from participatory activities

## 4.1. Survey results

We received eleven answers to the survey (9 men and 2 women) from different age ranges (1 ranging from 18 to 30 years, 6 from 31 to 45 years and 4 from 46 to 65 years) and from five different cooperatives (Figure 1). Two cooperatives - Käsivarsi (two answers) and Hammastunturi (one answer) - are located in the mountain area, whereas three cooperatives - Palojärvi (five answers), Sattasjärvi (two answers), and Oivanki (one answer) - are located in the forest area. For half of the respondents, reindeer herding was their main occupation, whereas the other half indicated that it was an activity where they occasionally help others, a hobby, or a way to get some revenues from tourism. Details on the results of the survey can be found in the Supplementary Information (Tables S1-S4, Figures S2-S4 and Text S2).

During summer, long hot periods can affect reindeer herding activities, as confirmed by seven survey respondents. At least one respondent indicated that this occurred in each of the last five years. The main impact of long hot periods is that calves suffer from heat and insects, especially during calf-marking, which forces the herders to reschedule their activities. In order to properly plan their activities, herders would need more information on the warm weather periods, their length, and maximum temperature and rainfall. According to the majority of answers, receiving this information one to two weeks in advance would be optimal, whereas one respondent indicated that this information would be useful two months in advance. Warmer growing seasons in summer was another situation identified by two respondents to have an impact on the harvesting work. Both respondents indicated that this situation allows them to finish harvesting earlier and devote more time to other activities. The respondents mentioned that knowing if the beginning of summer will be warmer than normal can be helpful to plan harvesting and that this information would be useful two weeks in advance.

In autumn, the respondents agreed that the weather condition that affected their herding activities was a warm autumn with a late appearance of soil frost and a late formation of snow and ice cover. Such a warm autumn usually brings a late rut that is not synchronised among reindeer (six respondents), difficulties in gathering and moving herds to the round-up sites (five respondents), decreased calf weight due to delayed round-ups (three respondents) and increased transportation expenses and working hours (two respondents). Therefore, the herders usually need to increase reindeer supervision and control and make sure to avoid weak ice, which affects travel conditions. Indeed, certain types of ice are better for transport, and lakes and rivers must have a thick ice layer to bear reindeer, humans and vehicles (Eira 2022). When the ice layer of river and lake sections is thick enough and covered in snow, it can hardly be distinguished from dry land (Krause, 2016). In this regard, survey respondents indicated that having information about temperature extremes, rainfall, and periods of warm weather would help them. Six respondents said they would like to receive this information two weeks in advance, but responses ranged from one week to three months.

In winter, the main condition affecting reindeer herding (ten respondents) is ice formation on the ground. The ice layers make it more difficult for the reindeer to find food, causing them to break loose to find higher availability of forage. Herders then need to give more food to the reindeer, resulting in higher expenses. Some herders (two) said they could cope by increasing the use of pastures, but most (six) needed to introduce supplementary feeding. Herders also needed to increase supervision and control of the reindeer, sometimes having to fetch them from further away. Another winter condition mentioned only by one respondent was winter storms with strong winds, which make it difficult for '*reindeer to travel against the wind*'. The respondent was from the Käsivarsi cooperative, located in the most North-Western part of the herding management area, which is closer to the ocean. These conditions limit the movement of animals to the grazing areas and the respondent indicated the need to introduce supplementary feeding to cope with these situations.

Regarding springtime, five herders, all from the cooperatives in the south of the surveyed area, said that cold events that occur suddenly after spring has already started, affected their herding activities. These situations are referred to as backwinter (*takatalvi*). As the calves are usually born in spring, they are still young and vulnerable to cold weather. The climate information that may be of use to them to adapt is the timing and the length of the cold weather period (four respondents). Temperature and rainfall information is also relevant. All the respondents needed the information one to three weeks in advance, with most (four respondents) stating two weeks in advance. Other climatic conditions identified in spring were early and late snowmelt, which impact the availability of fresh forage, reindeer movement and capacity to feed in nature. Three respondents from cooperatives located in the south of the surveyed area mentioned the relevance of information such as the duration of the warm weather period, maximum temperature and rainfall, which affect early snowmelt. Late snowmelt was mentioned by two respondents from cooperatives in the north of the surveyed area, who indicated that maximum and minimum temperatures and snowfall could be useful. Both for early and late snowmelt, the information would be useful if available two to four weeks in advance.

The current sources of weather and climate knowledge for reindeer herders are primarily websites (five), TV (four) and traditional knowledge based on observations of nature (four). The herders also share such information via WhatsApp groups (five) or text messages (three), as well as talking to each other about it in person (one).

When asked about the relative importance of climate change for reindeer herders, the survey showed that only two out of eleven respondents considered the long-term effects of climate change (i.e. beyond the next 20 years) as the main risk to their activities. However, this time horizon is not considered in the present study, limited to S2S time scales, since it would entail decisions related to long-term planning. Other more voted risks included land use conflicts (e.g. conflict with other economic activities such as renewable energy and forestry, indicated by nine respondents), costs of the herding activity (i.e. purchase of food and vaccinations, five respondents), price of the reindeer meat (five respondents), lack of political support (five respondents) and workload (three respondents). This is in agreement with several studies showing that land use and predation are the priority worries of reindeer herders (Löf 2013, Kivinen 2015, Lépy et al. 2018, Österlin et al. 2023). Although the threat of predators was not initially included in our survey, the issue was raised in the discussions with reindeer herders, who mentioned the importance of protecting reindeer from predators and searching for the lost remains of reindeer to get compensation from the state (Turunen and Vuojala-Magga 2014). Long-term climate conditions have lower priority for reindeer herders than these other risks. However, there is a general agreement that the climate has become more extreme and uncertain in all seasons.

## 4.2. Decisions for adaptation stories

Using the inputs gathered from the survey and informal interviews, we identified different decisions to co-develop 'climate change adaptation stories'. In this work, adaptation stories are used as a way to illustrate and analyse various decision making points that reindeer herders encounter during the herding season. Apart from understanding the decision making context of herders, this allows us to jointly explore the best available climate information for anticipating situations of interest for herders in a flexible way. The decisions co-explored in this study relate to the anticipation of: harvest time and insect harassment (summer), round-ups and the mating process (autumn), supplementary feeding (winter), the release of reindeer for outdoor feeding (spring) and inopportune backwinter episodes (spring). Adaptation stories are summarised in Table 1.

**Table 1.** Climate change adaptation stories to co-explore pilot climate services for reindeer husbandry. The climate service product indicated in the table corresponds to the option proposed and discussed during the roundtable.

| Description | Climate service product | Observed years | Impact on decision making and/or coping strategy |
|---|---|---|---|
| **Summer: Anticipating harvest time** | | | |
| Hay making in mid-summer has become a normal activity of reindeer herders. In the southern herding associations in Finnish Lapland, supplementary/corral feeding in winter is necessary to compensate for the loss of natural feeding. When it's very hot at the beginning of summer, the growing cycle of hay is accelerated | Average temperature in May and/or June (2 weeks in advance) | 2018 2020 2022 | In warmer than normal summers, reindeer herders finish the hay harvesting work already in June, which is earlier than normal. This leaves free time to devote to other activities |
| **Summer: Anticipating insect harassment** | | | |
| Rising summer temperatures and increasing rainfall increase the nuisance caused by blood-sucking insects and may allow new parasites such as deer flies to | Average temperature and precipitation in June and/or | 2018 (most remembered year) 2019 2020 | In hotter and wetter than normal summers, herders can reschedule calf marking |
| spread to the reindeer herding region. Insects are especially annoying during calf marking | July (2-3 weeks in advance) | 2021 2022 | |

**Autumn: Anticipating roundups and the mating process**

| | | | |
|---|---|---|---|
| During warm autumns, there is a late appearance of soil frost and a late formation of snow and ice cover. This makes it more difficult to gather and move herds to the round-up sites and calves weight is lower because there is less food in the nature | Autumn minimum temperature (2 weeks in advance) | 2019 2020 2021 2022 | In warmer than normal autumns, herders may have an increased supervision of reindeer herds and avoidance of weak ice |

**Winter: Anticipating winter supplementary feeding**

| | | | |
|---|---|---|---|
| If there is ice formation on the ground or in the snowpack in winter, this makes it more difficult for reindeer to feed, since they need to spend more energy digging and moving, which can affect animal condition. Herds 'break loose' to find forage, and there is an increase in transportation expenses and working hours | No climate product explored during the roundtable, since these phenomena occur at shorter time scales (hours, days) than the S2S ones explored in this work | 2019 2020 (most remembered year) 2021 | If fast temperature changes occur around zero degrees together with rainfall, signalling the potential formation of ice, reindeer herders can react by increasing the supervision of the herd and introducing more supplementary feeding (e.g. with lichen, hay, grass silage or pellets). While many herders buy the feed, others grow hay on their own fields if environmental conditions allow |

**Spring: Anticipating the release of reindeer for outdoor feeding**

| | | | |
|---|---|---|---|
| Early snowmelt in spring allows herders to release reindeer from fences earlier to feed in the forest. These conditions are especially favourable for lactating reindeer and the new-born calves | Average temperature in April (2-4 weeks in advance) | Not indicated (but in 2021 early snowmelt was observed) | If April is going to be warmer than normal, the snow will melt faster and reindeer herders may release reindeer earlier. Food can be found in the forest and this means that less food needs to be either cultivated or bought, allowing cost savings |

**Spring: Anticipating inopportune backwinter** *(takatalvi)*

| | | | |
|---|---|---|---|
| Backwinter is a situation when it is very cold in spring, when the calves are just born, and consequently it can affect the animal's survival. In addition, cold springs with heavy rains are considered harmful to young | Average weekly temperature in April and/or May (2-3 weeks in advance) | 2004 (most remembered year) 2014 | If it is known that back winter is coming, reindeer are not let out of the fences into the forest, but are kept for longer in the fences to be fed. |
| calves since their fur is not developed to regulate heat in the same way as an adult reindeer | | | |

## 4.3. Roundtable results

The climate change adaptation stories presented in Table 1 were discussed during the roundtable with reindeer herders. This allowed us to complete the table with additional details regarding the impact of the weather conditions on herders' decision making and coping strategies. It also allowed us to confirm whether the proposed climate service products were appropriate to advise on the identified decision.

The roundtable revealed that climate plays an essential role in the daily decisions of reindeer herders. For instance, rainy weather for many consecutive days, as well as large amounts of rain, can hinder herders in summer when preparing for the harvest (and also in autumn). Therefore, having such information a few weeks in advance can help them plan their activities during rain-free days. Besides rainfall, freezing and melting conditions during the start and end of winter are important. Information on temperature extremes is particularly useful in autumn, whereas information on temperature anomalies is welcome all year round. The herders would like to have more information on the arrival of spring, so a climate forecast for the month of May would be useful. In cases when a backwinter event occurs in the spring, information about its severity and duration is vital for the survival and development of young reindeer, as a week-long backwinter is a danger to young reindeer whose fur has not yet developed to regulate body temperature in the same way as an adult reindeer.

Other aspects mentioned were that given exogenous factors like the conflicts related to land use, the pressure on pastures, and the situation in other food supply regions (e.g. due to the war in Ukraine), shorter winter periods can increase food availability for reindeer, making animals less dependent on supplementary feeding.

## 5. Adaptation stories: complementing reindeer herders' autonomous adaptation with climate services

Situations that impact reindeer herders' activities have been identified for all seasons in Table 1. However, in this section we focus only on two adaptation stories to illustrate the application of seasonal and sub-seasonal climate predictions respectively: anticipating harvest time (section 5.1) and anticipating inopportune backwinter (section 5.2). These two situations were selected because they allow the investigation of the two different prediction horizons. Besides, specific years when such events had occurred in the past were available from the survey, which is crucial to check whether herders' perceptions are reflected in the observations. Finally, these stories were also useful to illustrate situations for which future predictions have and do not have enough quality to support herders' decision making, which provided a useful comparison for capacity building.

### 5.1. Anticipating harvest time

According to reindeer herders, 2018, 2020 and 2022 were years when harvesting finished in June, which is earlier than normal. In this section, we present the results for 2018 and 2020, since the gridded dataset used from the Finnish Meteorological Institute, covers the period 1961-2021, and therefore we do not have data for 2022. In 2018, the mean air temperature at the surface was above normal conditions for four consecutive weeks, starting in the week of the 7th of May (Figure 2, left). In 2020, the mean air temperature was also above normal conditions for five consecutive weeks starting on the 25th of May (Figure 2, right). Since these events show a duration of approximately one month, seasonal climate predictions, which aim to anticipate the climatic conditions for the next months and seasons, were selected for this particular application.

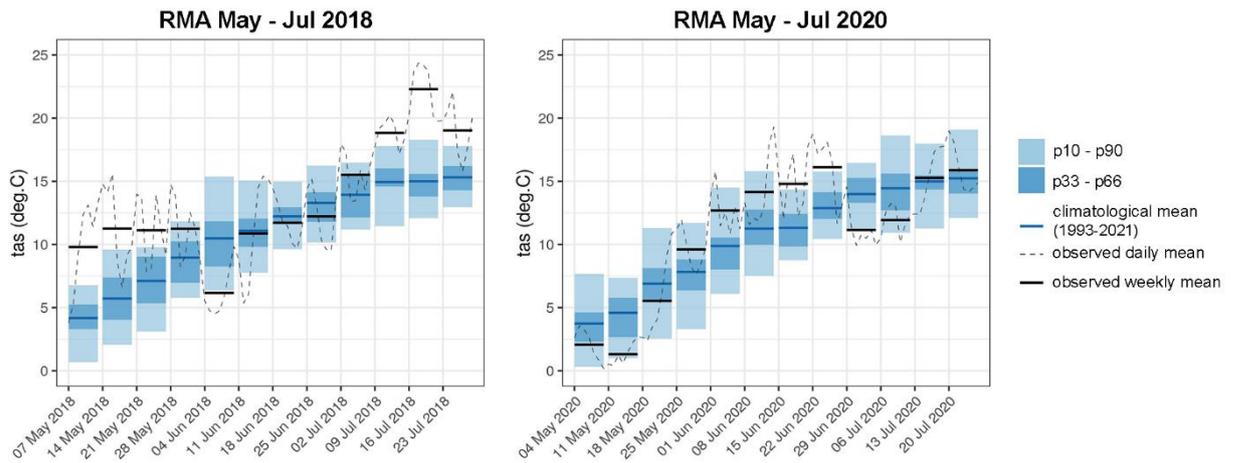

**Figure 2:** Temporal evolution of daily and weekly mean surface air temperature from May to July in years 2018 (left) and 2020 (right) compared to the climatological mean and percentiles p10, p33, p66 and p90 in the reindeer management area (RMA) in northern Finland.

While the event of May 2018 was an extreme event for which temperatures were above the 90th percentile (p90) in all the herding municipalities, June 2020 was also extreme in most of the region, but the three northernmost municipalities showed above normal temperatures and did not reach the upper extreme (Figure 3).

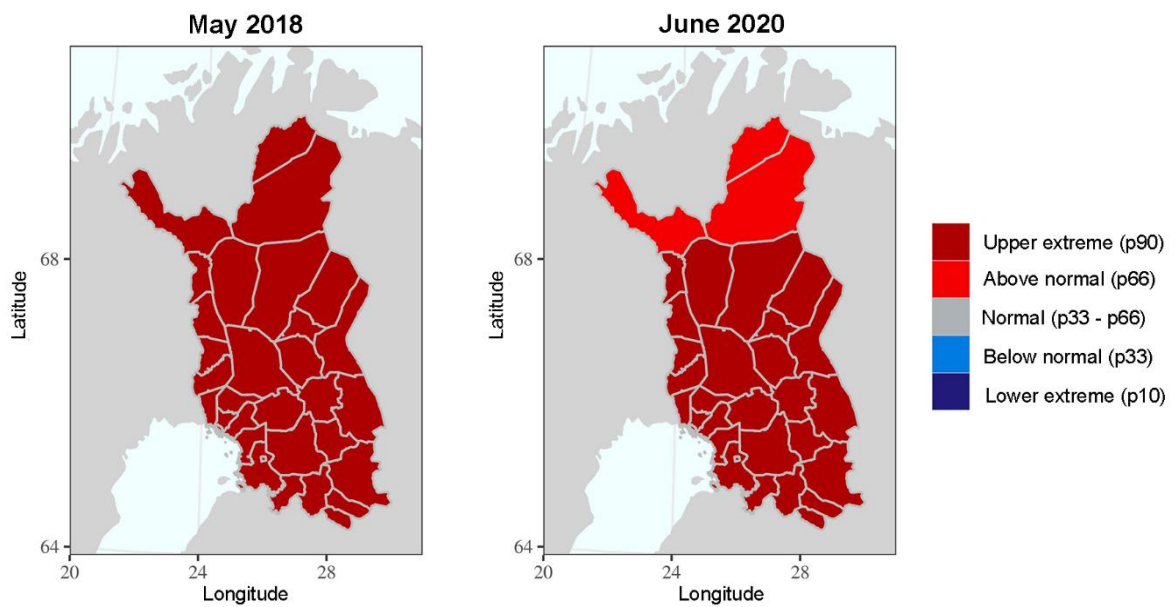

**Figure 3:** Spatial distribution of the observed categories for the monthly mean air temperatures in May 2018 (left) and June 2020 (right) in the herding municipalities in northern Finland.

### *What was predicted for summer 2018 and 2020?*

The seasonal prediction (SEAS5) for both target months (May and June) was explored to check the predictability of the event up to three months in advance. For the target month of May, the fRPSS for the three forecast months (predictions obtained in February, March and April) was positive (Figure 4), which indicates that seasonal predictions have enough quality to support decision making. However, for the June target month, the fRPSS was negative for all three forecast times (predictions obtained in March, April and May), meaning that, in this case, the climatology may provide a better estimate of future conditions than the seasonal prediction. In both cases, the predictions obtained three and two months prior to the event captured the most probable tercile category. Thus, in February and March 2018, a warmer than normal May was predicted with a probability of 45% and 38% respectively, whereas in March and April 2020, the probability that June was warmer than normal were 43% and 40% respectively. However, one month in advance, the prediction assigned a higher probability to the below normal category for both target months.

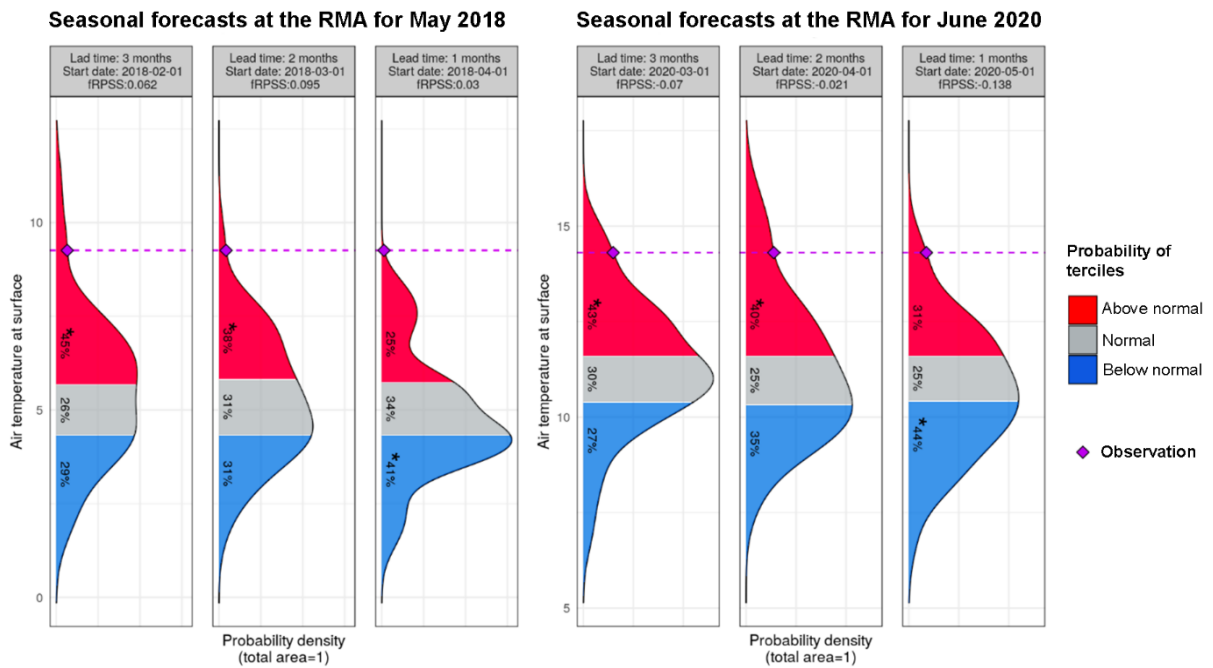

**Figure 4:** Post-processed seasonal (SEAS5) predictions for the monthly mean air temperature in May 2018 (left) and June 2020 (right) in the reindeer management area (RMA) in northern Finland.

Observed values are shown with a pink diamond, and the most probable tercile is indicated with an asterisk.

Reindeer herders recalled having earlier than normal harvests in the years 2018 and 2020. However, warmer periods were recorded in different months in these two years, according to observations, reflecting the complexity of co-developing tailored climate services. In this case, the definition of the climate adaptation story unveils two different climate service products from the service provider's point of view: seasonal prediction of May and seasonal prediction of June mean air temperature. This shows that the predictions for both the months of May and June may be informative for reindeer herders to know if an earlier than normal harvest is expected.

## 5.2. Anticipating inopportune backwinter

According to reindeer herders, 2004 and 2014 were two years with backwinter, involving periods of colder than normal conditions between the months of April and June. In 2004, the mean air temperature was below normal conditions for the week starting on the 19th of April and the week starting on the 10th of May (Figure 5, left). In 2014, the mean air temperature was below normal conditions for two consecutive weeks, beginning on the 28th of April (Figure 5, right). Given that the duration of the identified events ranged between one and two weeks, seasonal predictions were not able to capture them. Therefore, sub-seasonal predictions are better suited for this application because they can anticipate events occurring in the next weeks. For illustration purposes, here we only show the case of 2004 backwinter, since it corresponds to the year that herders remembered the most according to the survey results (Table S4 in the Supporting Information).

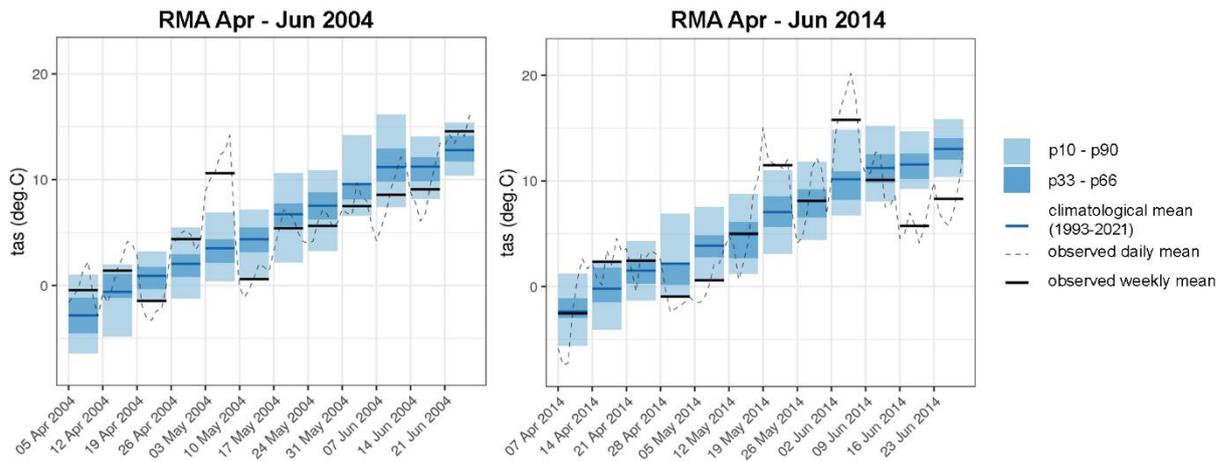

**Figure 5:** Temporal evolution of weekly mean surface air temperature from April to June in years 2004 (left) and 2014 (right) compared to the climatological mean and percentiles p10, p33, p66, and p90 in the reindeer management area (RMA) in northern Finland.

The spatial distribution of mean air temperature across the different municipalities shows that while the week starting on the 19th of April 2004 was a cold extreme event for the eastern municipalities of the reindeer management area (Figure 6, left), the mean temperature was below normal for the western part of the region, and normal in the municipalities located closer to the Gulf of Bothnia. For the week starting on the 10th of May 2004, the observed mean air surface temperature was below normal in all the municipalities of the herding region (Figure 6, right).

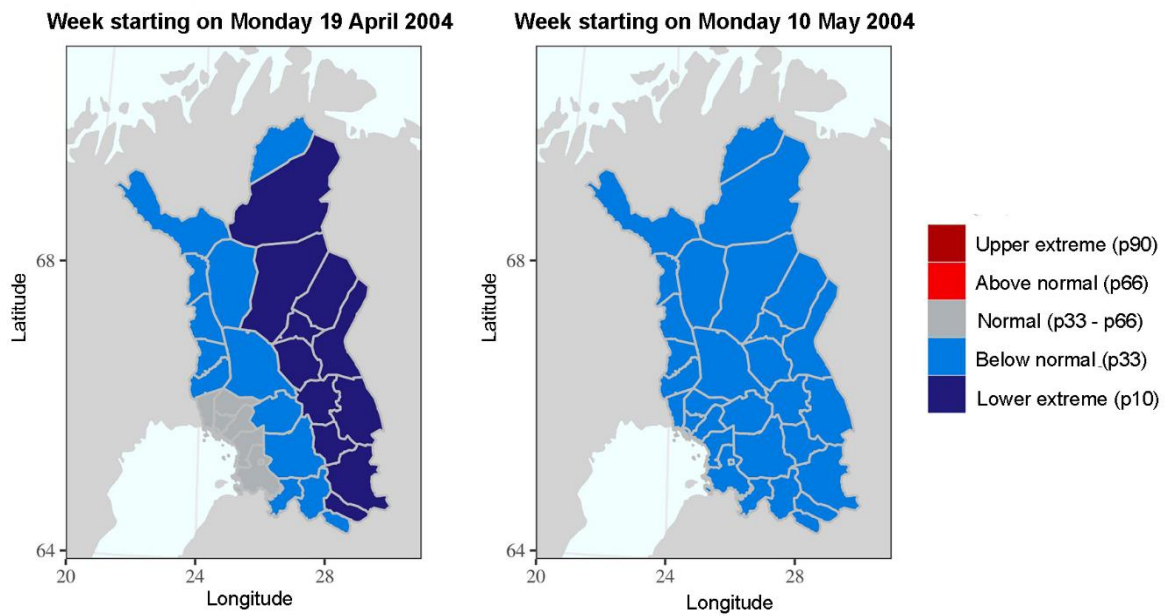

**Figure 6:** Spatial distribution of the observed categories for the weekly mean air temperatures on the week starting the 19th of April 2004 (left) and on the 10th of May 2004 (right) in the herding municipalities in northern Finland.

### *What was predicted for spring 2004?*

The sub-seasonal (NCEP CFSv2) predictions for one to four weeks prior to the event show positive fRPSS in all the cases, indicating the suitability of adopting the predictions to support reindeer herders' decision making (Figure 7). However, for the specific target weeks, only in one case was the forecast system able to capture the most probable tercile category for temperature over the whole reindeer management area (i.e. for the week starting the 10th of May, which is one week in advance of the occurrence of backwinter). In this case, a week earlier, the probability to have lower than normal temperatures that week of the 10t of May was 65%, with a 26% probability of having extreme values of temperature.

In this case, the prediction of the average weekly temperature in April and May is useful to anticipate whether colder than normal conditions lasting for at least 1 week are expected during this period. Knowing that colder than normal conditions lasting for at least one week that month would have been useful for reindeer herders to decide for how long to keep reindeer in fences to be fed, avoiding putting the survival of young calves at stake.

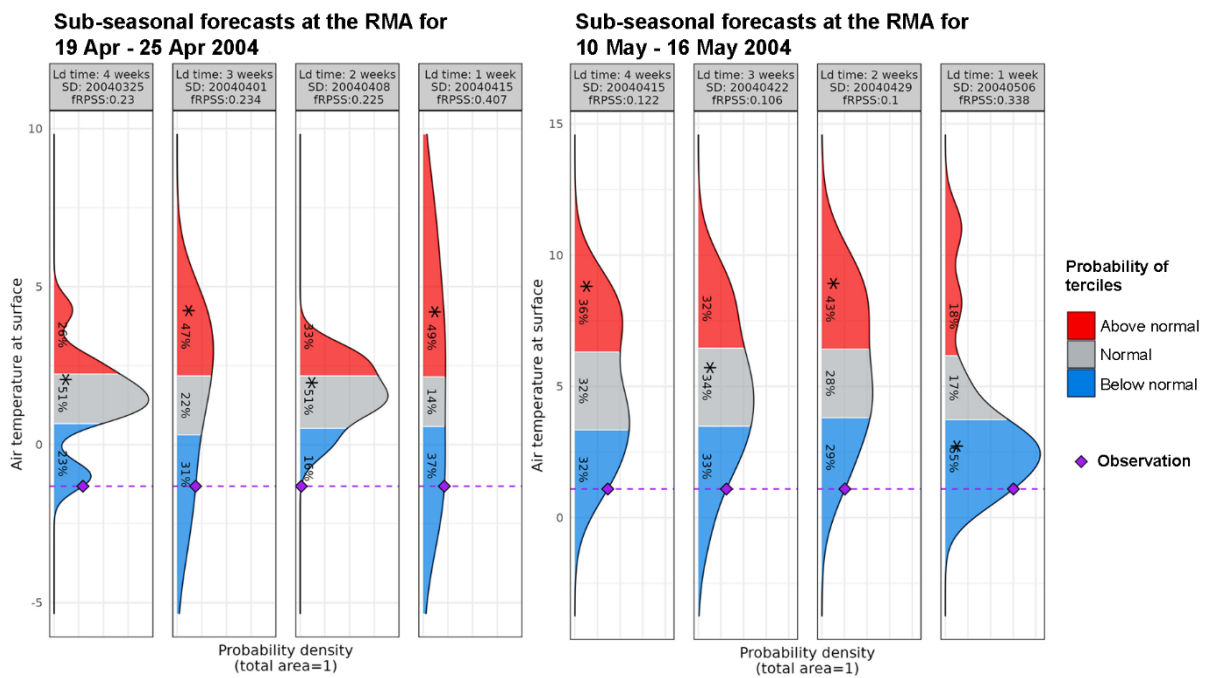

**Figure 7:** Post-processed sub-seasonal (NCEP CFSv2) predictions for the weekly mean air temperature for the target week starting the 19th of April 2004 (left) and the 10th of May 2004 (right) in the reindeer management area (RMA) in northern Finland. Observed values are shown with a pink diamond, and the most probable tercile is indicated with an asterisk.

For both selected climate adaptation stories (anticipation of harvest and backwinter), we have identified S2S predictions with enough quality to be used in decision making (i.e. with positive fRPSS values). However, attention should be paid to the interpretation of these results, so that users appropriately understand their benefits. Hence, it is important to understand that the quality of a prediction depends on a combination of different factors, including the climate variable of interest (temperature is normally easier to predict than precipitation or wind), the geographical region (the tropics tend to have a better predictability than higher latitudes), the month or season of the year (the quality of the prediction tends to be better in summer than in winter) as well as how far in advance the prediction is issued (in general, the more in advance we want to predict a situation, the lower the quality of the prediction). This means that not all the situations can be predicted with the same level of certainty, but we can find windows of opportunity or, in other words, situations in which the prediction might benefit reindeer herders.

Taking the case of backwinter as an example, the prediction for year 2004 (where the first colder than normal week could not be predicted, and the second week was predicted only a week in advance) may discourage reindeer herders from adopting such predictions in their decision making strategies if they only look at this particular year. However, it is important to understand that when assessing the benefits of climate predictions, we cannot base their performance on individual years. We need to be aware that the predicted most likely category can miss the observation in some particular years. However, in areas where the prediction is skillful (i.e. positive fRPSS values), using the prediction will always be better than using the climatology. There is therefore a need to move from a short- to a long-term approach, since the benefits from adopting climate predictions can only be perceived in the long term (Terrado et al. 2019).

## 6. Discussion and conclusions

To shift it from being *available* to being *used*, climate data must be communicated and delivered in a suitable way to be understood, accepted, and incorporated in the already existing practices of potential users. This work has set the basis for a successful collaborative mode of engagement and research in the co-production of climate service products that can support the decisions of Arctic reindeer herders. Using climate change adaptation stories connecting herders' decision making with climate conditions, we have been able to relate individuals' experiences and understanding of climate change to adaptation actions. For that, having a person that acts as liaison and who knows the stakeholders personally has been crucial, as it takes time to build trust and common understanding. The roundtable discussion with reindeer herders highlighted the importance of language as a part of the provision of a climate service, especially when approaching different cultures. Language has been identified as a barrier that can hinder the efficient communication among participants (Latola et al. 2020). Translation efforts cannot be underestimated and are necessary for an effective interaction, since being able to use one's own words reflects the identity of participants and determines whether they feel represented and able to voice their opinions (Kovach 2010).

The main challenge for the co-development of the climate change adaptation stories has been translating weather and climate situations into suitable climate service products that support anticipatory and planned adaptation. For this, we have co-explored S2S climate predictions for two particular situations related to the prediction of harvesting time and the occurrence of inopportune backwinter episodes. The difficulties of defining a particular prediction product can derive from the complexity of the situation, which may differ year by year (i.e. the prediction of early harvesting identified a different target month in 2018 and 2020), and also from the discrepancies between herder perceptions and the climate data, which have been related to the biases in human memory and the difficulty to separate possible climate change from interannual variability (Rasmus et al. 2020).

This study focuses on two situations (harvest time and inopportune backwinter) for illustrative purposes, highlighting that additional herding decisions may benefit from predictions at sub-seasonal and seasonal time scales. Therefore, having information at these time scales in advance may allow reindeer herders to manage their day-to-day activities and plan future tasks more efficiently, potentially improving their production and time management, and reducing the use of inputs and their associated costs (e.g. food, vaccines, etc.). We show that climate predictions may be useful for situations in which they have enough quality for decision making, acknowledging that their quality may be more limited in other situations, depending on the variable of interest, the geographical region, the predicted month or season and how far in advance the prediction is needed. The next steps for the uptake of this proof-of-concept climate service will depend on the ability of scientists to communicate its added value in an effective manner to the broader herding community as well as how keen herders are to integrate this new knowledge in their decision-making.

Additional efforts should include a co-evaluation of this proof-of-concept climate service to assess its usefulness in real world settings, assess which climate services are more suitable for upscaling, and better understand how this knowledge can complement the traditional knowledge and autonomous adaptation practices of reindeer herders, which relies on both traditional practices and new techniques that have appeared more recently to deal with variable weather (Bojovic et al. 2015), encompassing climate services as the ones investigated in this work. The co-evaluation should also consider the testing of culturally appropriate forms of visualisation and communication of observed and predicted changes in Arctic climate to facilitate the uptake of climate information by reindeer herders. In addition, it would be interesting to co-explore other climate forecast systems as well as complex processes (e.g. basal ice formation) and ways to combine information from different data sources (e.g. mosquito life cycle). The connection between climate change and health also calls for an integration of co-creation and OneHealth approaches that consider the close connection between humans, animals and the environment.

We believe that the evidence gathered in this study, when tested under future scenarios, can pave the way towards an operational climate service for reindeer herders. If appropriately integrated with the traditional knowledge of Arctic communities, this new co-produced knowledge will help increase the capacity of these communities to deal with climate change and variability. Overall, this will contribute to increasing the resilience of Polar regions in the face of climate change, offering opportunities for adaptation while supporting the sustainability of the culture and traditional practices existing in the Arctic region.

**Acknowledgements**

This work has received funding from EU-PolarNet 2 via the European Union's Horizon 2020 research and innovation programme under grant agreement No 101003766. We acknowledge the reindeer herders from the Käsivarsi, Hammastunturi, Palojärvi, Sattasniemi and Oivanki cooperatives that participated in the survey and the herders from the cooperatives of Palojärvi and Näkkälä that kindly took part in the round table. We also thank Pierre-Antoine Bretonnière from BSC for downloading the climate data used in this work.